\documentstyle[12pt,epsfig,rotate,aaspp4]{article}

\newcommand\be{\begin{equation}}
\newcommand\ee{\end{equation}}

\newcommand{\mcommand}[1]{\ifmmode #1\else $#1$\fi}
\newcommand\Rtel{\mcommand{R_{\rm tel}}}
\newcommand\RBOSS{\mcommand{R_{\rm BOSS}}}
\newcommand\DBOSS{\mcommand{D_{\rm BOSS}}}
\newcommand\Dsat{\mcommand{D_{\rm sat}}}
\newcommand\Umask{\mcommand{U_{\rm mask}}}
\newcommand\rhosource{\mcommand{\rho_{\rm s}}}
\newcommand\phisource{\mcommand{\phi_{\rm s}}}
\def\sci#1{\mcommand{\times 10^{#1}}}

\def\unit#1{\mcommand{\;{\rm #1}}}
\newcommand{\kelvin}{\unit{K}}
\newcommand\meter{\unit{m}}
\newcommand\second{\unit{s}}
\newcommand\as{\unit{as}}
\newcommand\mas{\unit{mas}}
\newcommand\parsec{\unit{pc}}
\newcommand\kpc{\unit{kpc}}
\newcommand\yr{\unit{yr}}
\newcommand\km{\unit{km}}
\newcommand\kmps{\unit{km\;s^{-1}}}
\newcommand\pmsqs{\unit{m^{-2}\;s^{-1}}}
\def\micron{\unit{\mu m}}
\newcommand\gram{\unit{g}}
\newcommand\kg{\unit{kg}}
\newcommand\AU{\unit{AU}}

\def\vec#1{\mbox{\boldmath$#1$\unboldmath}}

\newcommand\etal{et al.}
\newcommand\etc{etc}

\newcommand\ltsim{\lesssim}
\newcommand\lsim{\lesssim}

\newcount\fixmecnt \fixmecnt=0
\def\fixme#1{\advance\fixmecnt by1\par\noindent{\sc FIXME 
\the\fixmecnt : #1}}
\def\endfixme{\par\noindent{\sc END FIXME \the\fixmecnt}}

\begin{document}

\rightline{\bf CWRU-P17-99}
\rightline{Submitted to: \em The Astrophysical Journal}

\title{The Big Occulting Steerable Satellite (BOSS)}
\author{Craig J. Copi\altaffilmark{1} and Glenn D. Starkman\altaffilmark{1,2}}
\authoremail{cjc5@po.cwru.edu}
\authoremail{gds6@po.cwru.edu}
\affil{
EMail: {\tt cjc5@po.cwru.edu} and {\tt gds6@po.cwru.edu}\\
BOSS home page: {\tt http://erebus.phys.cwru.edu/$\sim$boss/}}
\altaffiltext{1}{Department of Physics, Case Western Reserve
University, Cleveland, OH 44106-7079}
\altaffiltext{2}{Department of Astronomy, Case Western Reserve
University, Cleveland, OH 44106-7079}
\authoraddr{10900 Euclid Avenue\\ Cleveland, OH 44106-7079}

\begin{abstract}
Natural (such as lunar) occultations have long been used to study sources on
small angular scales, while coronographs have been used to study high contrast
sources.  We propose launching the Big Occulting Steerable Satellite
(BOSS), a large steerable occulting satellite to combine both
of these techniques.  BOSS will
have several advantages over standard occulting bodies.  
BOSS would block all but about $4\sci{-5}$
of the light at 1 micron in 
the region of interest around the star for planet detections.
Because the occultation occurs outside the telescope,
scattering inside the telescope does not degrade this performance.  
BOSS could be combined with 
a space telescope at the Earth-Sun L2 point 
to yield very long integration times, in excess of $3000\second$.
If placed in Earth orbit, 
integration times of $160$--$1600\second$ can be achieved 
from most major telescope sites for objects in over 90\% of the sky.
Applications for BOSS include  direct imaging of planets around nearby stars. 
Planets separated by as little as $0.1$--$0.25\as$ from the star they orbit
could be seen down to a relative intensity as little as $1\sci{-9}$ around a
magnitude 8 (or brighter) star.
Other applications include ultra-high resolution imaging of compound sources,
such as microlensed stars and quasars, down to a resolution as little as
$0.1\mas$.
\end{abstract} 

\keywords{planets -- gravitational lensing---occultations---space vehicles---stars:
low-mass, brown dwarfs}

\section{Introduction}

The search for planets around nearby stars is a major objective of modern
astronomy.  Recently, astronomers have discovered Jupiter-mass
planets orbiting nearby stars (see e.g., Mayor \& Queloz~1995; Butler \&
Marcy~1996; Cochran, \etal~1997; Marcy, \etal~1997;
Noyes, \etal~1997; Delfosse, \etal~1998; Marcy, \etal~1998; Delfosse,
\etal~1999).  They observed the periodic variation in the central
stars' velocities due to their motion around the center-of-mass of the
star-planet systems.  However, they were unable to image the planets
directly---the planets are too close to the stars that they orbit, so
diffraction in the telescope and atmospheric seeing cause the starlight to
wash out the planets.  This problem is generic to observations of compact,
compound sources in which the range of brightnesses of the individual
components is large.

On another forefront, many astrophysical systems have structure on
milli-arcsecond or sub-milli-arcsecond angular scales. 
Examples of this include image splitting of background stars in microlensing
events by massive compact halo objects (MACHOs), binary stars,
globular cluster cores, Galactic supernovae, the accretion disk around a
massive black hole at the center of the Milky Way, cores of nearby galaxies,
lensed images of galaxies, and distant galaxies and clusters all have
structure on the milli-arcsecond scale. Apparent angular displacements due to
parallax and proper motions are also on this angular scale for objects in the
nearest few kpc of the Galaxy.  The proper motion of a star at $10\kpc$, with
a velocity relative to the Sun of $200\kmps$, is approximately $5\mas\yr^{-1}$.

The separation of dim sources from nearby bright ones, and the resolution of
images at milli-arcsecond scales in the UV, optical, and near IR are limited
principally by diffraction, $\Delta\theta\geq\theta_{\rm diff}\simeq
1.2\lambda/D$.
Here $\lambda$ is the wavelength of the observations and $D$ 
is the diameter of the telescope or the baseline of the interferometer.
On the ground the effects of seeing on resolution can be reduced 
using adaptive optics (AO)
to nearly the diffractive limit in the IR, not yet at shorter wavelengths.
However, the diffractive limit at $1\micron$ on a 10-m telescope is still
$25\mas$, too large to resolve structures like Galactic microlenses.
Moreover, while AO is typically very good at squeezing the central core of the
point spread function (PSF) to near the diffraction limit, it still leaves the
halo of the PSF\@.  Getting all the way to the brightness contrasts of interest
for planet searches ($>10^8$) strictly on the basis of AO, is considered
technologically challenging in the IR, and daunting at shorter wavelengths.

Atmospheric seeing can be eliminated entirely by observing from space,
though scattering inside the telescope and diffraction persist.
These combined effects make it difficult to observe 
most expected planetary systems from the existing Hubble Space Telescope,
and  even for the Next Generation Space Telescope (NGST).
Greater resolution can be obtained using interferometry, 
and plans exist to do both ground-based and space-based IR 
and optical interferometry.  
The Space Interferometry Mission, a $10\meter$ baseline interferometer
expects to do astrometry at the microarcsecond level, and so
directly detect the  wobble in a  star due to an orbiting Jovian planet,
but will not separate the light of the planet from that of the star.
A space-based interferometer the same size as BOSS, 
such as the proposed Terrestrial Planet Finder, 
would be superior to BOSS for high resolution imaging of complex sources, 
though for high-contrast observations 
or resolution of simple multi-component sources,
it too might benefit if combined with a scaled up version of BOSS.
Such very-large space-base interferometers will be
more challenging technologically, more costly, and require more time
to build than BOSS.

Lunar occultations have long been used to resolve small-scale angular
structure (see e.g., Han, Narayanan, \& Gould~1996; Simon \etal~1996;
Mason~1996; Richichi 
\etal~1996; Cernicharo \etal~1994; Adams \etal~1988).  
As the Moon orbits the Earth, 
it sweeps eastward across the night sky.  
If it occults another source,
by monitoring the light-curve from that source, 
one can deduce the integrals of the source surface-brightness 
perpendicular to the apparent lunar velocity as a function of angular position 
parallel to the lunar velocity.  For a binary source this means that
one can measure the separation of the source components projected along the
moon's velocity vector, and the relative intensity of the two sources.

There are however several problems with using the Moon this way:

\noindent
(1) The lunar orbit is fixed; sources cannot be scheduled for occultation.  
{}From the ground, sources at ecliptic latitudes greater than about $5^\circ$ 
 are never occulted; from space, the fraction of the sky which is occulted
is similarly small.

\noindent
(2) The apparent angular velocity of the Moon 
with respect to the background stars ($0.55\as\second^{-1}$) 
is large and fixed by the orbital speed of the Moon.

\noindent
(3) The Moon is bright despite its low albedo.
Even the dark side is visible to the naked eye.

To avoid these difficulties, we are investigating the scientific
merits of launching a large occulting satellite.  The Big Occulting
Steerable Satellite (BOSS) will consist of a 
large occulting mask.  Patch geometry and transmission function
can be optimized for either high-resolution image reconstruction or
separation of bright from dim sources.  
In this paper we will concentrate on the  latter.
The configuration that we will consider is a  square structure with
a circularly symmetric transmission function that rises smoothly from 
zero (opaque) at the center to maximally transmissive where the
inscribed circle osculates the edge of the square.
The size of the inscribed circle, as will be made clear below,
needs to be large, approximately $35\meter$ in radius.  Similar ideas have 
been proposed independently in the past (see eg.\ Schneider~1995).  However
here we more 
thoroughly discuss the optical and mechanical properties as well as the
capabilities and challenges of such
a satellite.  Furthermore we have developed a configuration for BOSS that
would clearly contribute to several outstanding astronomical challenges, and
that is
viable for launch in the next 5 years.

Given BOSS's size, one needs to do whatever possible to minimize its mass.
The most likely approach is to fashion BOSS from a thin film supported
by  a framework of inflatable or deployable struts.  
Appropriate films with surface densities of $\sigma \simeq 10-20\gram\meter^{-2}$
are available, and the technology
for deploying large structures composed of them reliably is
under active development.

In the near term there are two possible configurations for the satellite.
The first is to deploy BOSS at L2, the second Lagrangian point
of the Earth-Sun system $0.01\AU$ past the Earth along the Sun-Earth line, 
in conjunction with a
large space telescope such as NGST (8-m class telescope).  The telescope and
BOSS would 
orbit the L2 point, with BOSS being steered to cause the 
desired occultations.  We term this the space-space configuration.  The
alternative is to place BOSS in a high apogee eccentric Earth orbit,
with observations done using ground-based telescopes.
We term this the space-ground configuration.

The space-space configuration has certain definite advantages.
Problems of reflected sunlight entering the telescope
would be considerably reduced
because of greater  flexibility in the 
telescope-satellite-source-sun-earth geometry;
orbital dynamics would likely be simpler, more stable and more predictable.
By using a combination of ion engines and 
sailing in the solar radiation pressure, 
one should also be able to obtain very long integration times.
A main drawback is that one would incur the
the added expense of a space telescope, 
however current plans are for NGST to be located in a halo orbit around L2.

The space-ground has advantages in cost and risk 
since no astronomical instruments are deployed in space.  
Moreover, the satellite could be used from any telescope, 
including those of amateur astronomers, 
by making satellite coordinates publicly available over the Internet.
Furthermore, since one is using ground-based telescopes,
they can be quite large, and moreover
any improvements in telescope or detector technology,
such as advancements in adaptive optics systems,
can be immediately used in conjunction with the satellite.
Aside from observing issues such as atmospheric seeing and
terrestrial thermal backgrounds,
the major complications in the space-ground configuration
are likely to be 
preventing excessive sunlight from being reflected into the telescope,
and ensuring adequate stability and predictability of the orbital dynamics 
given solar radiation pressure, magnetospheric and atmospheric drag, 
and gravity gradients.
Neither of these issues has yet been fully addressed,
though a brief discussion of the reflection issue is included below.
In this work we will focus on the space-space configuration.

Although solar sailing will likely be important in either configuration,
the satellite will undoubtedly not be able to rely entirely 
on solar sailing for maneuverability and ion engines will be required.
As will be discussed below, 
the requirements on the propulsion systems are not unreasonable,
and can probably be met with currently available systems.

\section{Location and Size of BOSS}

Before directly addressing BOSS's scientific capabilities we must address
some important logistical questions: 
Where should the observer be located?  Where should BOSS be located?
How big does BOSS have to be?  
These are, of course, related questions.  
In this paper, we will address them in the context of the space-space
configuration; similar questions will arise in the space-ground mode.

\subsection{Locating BOSS}

The criteria determining where BOSS should be located are: (1) longest
possible duration occultations, (2) easiest possible multiple target
acquisition, (3) highest possible duty cycle, and (4) lowest possible near-IR
backgrounds.

The first objective---long duration occultations---is 
both the most important and the most difficult to accommodate. 
In the occultation of a bright star to allow
the observation of an associated planet, 
the brightness contrast between the sources is usually very large, 
typically $> 10^8$.  In the geometric optic limit,
the satellite completely blocks the bright source during occultations.
However, because of diffraction by the satellite, some of the light from the
bright source reaches the telescope (see the discussion in section
\ref{sect:diffract}).  
Of course, the  larger the diameter of the occulter,
the greater the fraction of light which is blocked;
however, the  further away the planet must be from the 
occulted star not to itself be occulted.  
Thus there is a competition between the need to block
as much stellar light as possible and the desire
to detect planets as near as possible to the occulted star.
The matter is complicated by the fact that 
for a fixed solid angle subtended by the occulter, 
the fraction of light which arrives at the telescope
is a decreasing function of the  telescope-occulter separation.
Consequently, one would like to place the occulter as far away
from  the telescope as possible, while making the occulter
as large as possible.    Current technology for constructing
and deploying large, light structures in space suggests
that an occulter size of approximately $70\meter$ is 
achievable.   The relevant angular separation for planets
around nearby stars is approximately $1 \AU/10\parsec = 0.1\as$.
Thus ideally one would like a $70\meter$ diameter BOSS located at
a distance of order $10^4\hbox{--}10^5\km$.

Since the angular size of the satellite is chosen to be comparable to
the angular separation of the bright and dim sources,
the time over which one can integrate  on the dim source
is approximately the time during which the satellite  occults the bright 
source---essentially the time one can keep the satellite from
crossing its own length.
This must be long enough to allow the dim source to be detected 
above both the associated bright source and any background.  

We have identified two principle options for making long duration occultations,
one for a satellite in orbit around the Earth occulting ground based telescopes
(the space-ground configuration),
one for a space based satellite occulting a space based telescope
(the space-space configuration).

\subsection{Space-Ground Configuration}

Although we will not focus on the space-ground configuration we note that it
is possible to construct orbits with long integration times over most of the
sky for most telescopes at tropical sites (such as Keck).  
This is done by placing BOSS
in a high apogee orbit where the instantaneous tangential velocity at apogee
is equal to the rotational speed of the Earth (and hence the telescope).  Thus 
at apogee BOSS will appear to hang, or even undergo retrograde motion, over
the star.  For sources over 95\% of the sky orbits with 4 or 5 day periods can
be constructed that lead to $100$--$2300$ second occultations.

\subsection{Space-Space Configuration}

A number of factors persuade us to consider more
carefully the possibility of a space-space configuration
for the BOSS.  First, although long
occultations are possible in a ground-space configuration,
very careful orbit selection 
is necessary---one must insert the satellite
at precisely the  right place, with precisely the correct
velocity at precisely the right time in order to have
the velocity of BOSS cancel the velocity of the  telescope.  
Second,  the satellite period is of order days and only
one long occultation occurs per period, so the duty cycle is small.  
Third, changing from one target star to another 
is a complicated exercise in orbital dynamics, 
which can be costly in terms of both time and propellant.  
Fourth, space based telescopes are not subject to  atmospheric seeing,
nor to the same background levels as ground-based telescopes
in the infrared wavelengths most interesting to planet discovery.
Finally, space-based optical-IR telescopes to follow the Hubble Space
Telescope, such as NGST, are being planned.

The Lagrangian points of either the Earth orbit or any  other
planetary orbit hold particular attraction for
the deployment of a BOSS in conjunction with a space telescope.  This is
because the 
magnitude of the local effective gravity is   quite small,
vanishing exactly at the Lagrange point (by definition).
The unstable L2 point holds extra attraction because
it  is kept relatively clean  of destructive debris.
Indeed, NGST is likely to be deployed in a halo orbit around L2.
This point is located along the line from the Sun to the Earth
approximately $1.5\times10^6\km$ beyond the Earth.
Deployment of BOSS in conjunction with a  space telescope
at or near L2 is thus quite  attractive.  

At the L2 point the gravity due to the Earth and the Sun
add up exactly to give an orbital period for a circular orbit
of 1 terrestrial year.   Around the L2 point, there are elliptical orbits
in the ecliptic plane, and oscillatory vertical orbits perpendicular
to the plane all of period approximately $1/2$ year.
In addition, there are unstable modes in the  ecliptic of similar
time constant. These periods/time-constants are independent
of distance from L2, for distances $\lsim 10^4\km$.
At a distance $d$ from the L2 point the local gravity is therefore
approximately $a \simeq 4\pi^2d/\tau^2$ where 
$\tau \simeq 150\hbox{--}200$ days.
For $d$ in the range of  $10^{(4\hbox{--}5)}$km, this gives
$a\simeq 2\times 10^{-(5\hbox{--}6)}\meter\second^2$. This means that once BOSS
is positioned to occult a star it takes at least
${\cal O}\left((1\hbox{--}3)\times10^3\right)$ seconds
for BOSS to drift $10\meter$ off-center, with no corrective maneuvers.
Typical  orbital velocities around the L2 point are also relatively low:
$v\simeq 2\pi d/\tau$, which is $(4\hbox{--}40)\meter\second^{-1}$ 
for $d=10^{(4\hbox{--}5)}\km$,
making it relatively easy and inexpensive to reposition the BOSS
and station-keep with respect to the telescope.
For orbits out to approximately $10^5\km$, 
numerical simulations show similar stability
and low orbital velocities.

\subsection{Satellite Size}

Suppose we can reliably position BOSS in space
to a tolerance less than its linear dimensions.
The satellite must then be large enough to guarantee 
that scheduled occultations occur despite the uncertainty, $\Delta\delta$,
in the source location on the sky.
One of the big uncertainties that would face the BOSS project
is to what accuracy one can determine absolute star positions on the sky.  
Although Hipparcos determined the relative positions of stars in
individual fields to about $1\mas$, the global  fit to the 
positions of stars in the Hipparcos catalogue gives an accuracy
of  only about $\Delta\delta\ltsim0.1\as\simeq 5 \times10^{-7}\;$rad.
Moreover, since the baseline for proper motions was only
three years, 
as the epoch of the Hipparcos catalogue recedes into the past,
the accuracy of the star positions decreases.
This is an important effect  for
the stars which are the likely targets of searches for planets.
However,  the Hipparcos catalogue is likely to be supplemented
by a SIM catalogue before the BOSS launch date.  Since SIM
can determine relative positions within a field to about $1\;\mu$as,
we expect to have absolute angular positions  for the target stars
better than  $\Delta\delta\ltsim0.01\as\simeq5\times10^{-8}\;$rad.

We require that the BOSS be large enough to  subtend
a solid angle greater than the uncertainty in its desired position, 
$ r \Delta\delta$.
If $r\leq 2\times 10^8\meter$ and $\Delta \delta = 0.01\as$,
then we must require $x_\perp\geq 10\meter$.
For a diameter of $70\meter$, BOSS at $1\times 10^8\meter$ distance
would subtend about $0.14\as$.  
Thus if target stars have absolute positions determined to about 
$0.01\as$, then we  can accurately aim to occult the star near the center
of the satellite.  Realtime examination of the diffraction image of the star
would allow positioning well below $10\meter$.

Even if we can know where the satellite {\it is\/} very accurately,
can it be placed where it is needed when it is needed
with the appropriate velocity? 
That question is difficult
and is currently unanswered, but the answer is not obviously no.
Relevant effects include solar radiation pressure, 
lunar gravity, eccentricity of the Earth's orbit, \etc., all of which need to
be understood to address this question.

The second important effect influencing the choice for BOSS's size 
is the satellite's nulling efficiency---the fraction of light  from an
occulted source
that diffracts around the satellite decreases with the area of the satellite
(see section \ref{sect:diffract}).  
The higher the contrasts between bright and dim sources that one wants to study,
the larger the angular size that BOSS needs to be.  
On the other hand, the larger BOSS is, the further away on the sky
the dim source must be from the bright source to avoid getting occulted itself.
For high resolution imaging,  
there is less constraint since in the Fresnel
limit the diffraction limit for the satellite
is independent of the size of the satellite if
$\lambda < x^2/r \simeq 500\micron$.

We will use $\RBOSS=35\meter$ for BOSS in  this paper,
though further characterization of the performance BOSS as a
function of size is essential and ongoing. 
The optimum answer will depend on the exact
observing program and the nature of the objects to be occulted.

\subsection{Satellite Mass}

While there may be scientific pressures to make the
occulter as large as possible, there
are clear financial pressures to keep its mass as low as possible.
In particular, the cost of launching a satellite increases dramatically
with mass.  
Hence the maximum allowable satellite area
$\pi \RBOSS^2$ 
depends on the minimum film thickness one can tolerate,
the strength of any framework, etc\@.
Since skin depths of good conductors in optical and IR bands are
very much less than a micron, 
one can readily render the occulter as opaque as needed.
The minimum film thickness is then determined by structural integrity 
not optical necessity.
Appropriate films exist with surface density 
$\sigma\simeq10-20\gram\meter^{-2}$,
so  a single layer $70\meter$ diameter circle would have a mass of 
$40\hbox{--}80\kg$.

The rigidity of the satellite structure could be achieved 
by supporting the film using  inflatable, self-rigidizing struts.
Because the struts rigidize after inflation, micrometeoroid hits would not
affect the satellite structurally.  Micrometeoroid impacts on the film might,
however, affect the optical properties of BOSS\@.
Small holes, with diameters less than the wavelength at which 
observations are planned ($\lambda\simeq 1\;\mu{m}$), 
would not severely impact the performance of the occulter
until their areal density became quite large.
This is because the photons encountering such holes would
see them only as small changes in the effective transmission function of the
satellite, 
and because diffraction would spread any passing light ray into a  wide beam.
Much larger holes would be more problematic, and could degrade
performance significantly at even a very low level.
Attitude and orbit corrections must be 
sufficiently gentle so as not to damage the framework or tear the film.
Preliminary estimates show that struts and deployment system to support
a $35\meter$ radius occulter able to withstand accelerations of 1-2 g's,
would total another $100\hbox{--}200\kg$.
Also, one would like the amplitude of oscillations of the satellite to be small
so that sunlight is not reflected into the telescope.
Detailed analysis of the modes of oscillation, damping times, 
maximum allowable amplitude of oscillation, as well as the
permitted levels of film deterioration will also be required.

\subsection{BOSS Geometry}
\label{sect:geometry}

Supporting a circular film is difficult.  Putting a rigid, opaque ring around
the circumference of BOSS destroys the nulling achieved by using a radially
dependent transmission function unless the ring is extremely narrow. We
instead suggest imprinting a
circular transmission function on a transparent square film.  The square will
be supported by rigid, opaque struts along its diagonals.  Our model
configuration is thus made up of a $70\meter\times70\meter$ transparent square 
with a $35\meter$ radius, radially dependent, circular transmission function
inscribed.  It is supported by 20 inch ($0.5\meter$) diameter struts. 
The optical properties of this arrangement will be explored in
section~\ref{sect:planet}.

\section{Steerability}

To change the satellite orbit to occult a particular object requires imparting
an impulse:
$\Delta {\vec p} = m\Delta {\vec v},$
resulting in a change of velocity
\be
\frac{\Delta\vec v_{\rm sat}}{v_{\rm sat}} 
= \frac{\Delta m_{\rm propellant}{\vec v}_{\rm ejection}}{m_{\rm sat}v_{\rm
sat}} .
\ee

The solar radiation pressure will also play a significant role
in determining the satellite's motion.  This can either be viewed as
a problem or an opportunity.   The solar radiation pressure
in the vicinity of the Earth (and L2) is approximately
$P = 4\times10^{-6} \unit{N}\meter^{-2}$.  For a satellite of effective 
surface density $0.05\kg\meter^{-2}$, this translates into an acceleration
of $8\times10^{-5}\meter \second^{-2}$.  
This means that the satellite could cross a $30,000\km$ radius orbit
in 1--2 days, or a $100,000\km$ radius orbit in 2--4 days
using solar radiation pressure.  Of course, the solar radiation
pressure is pushing away from the sun, and L2 is unstable,
so the use of solar sailing is somewhat limited. 
However, understanding solar radiation pressure {\it will\/} be essential,
and utilizing it would be desirable.

Whatever use BOSS makes of solar sailing, it will clearly be necessary
to do some orbital adjustment using rockets.
The number and size of such adjustments will probably be the limiting factor
on the useful lifetime of the satellite.
If $N$ is the number of desired major rocket-driven orbital corrections,
then we must keep
$(\Delta m_{\rm propellant}/m_{\rm sat})\leq N^{-1}$,
where $\Delta m_{\rm propellant}$ is the typical mass of propellant 
expended per orbit reconfiguration.
We therefore need  
\begin{equation}
v_{\rm ejection} \geq N \Delta{\vec v_{\rm sat}} .
\end{equation}
As discussed above, satellite orbital velocities are  in the range 
$4\hbox{--}40\meter\second^{-1}$
for the orbital radii of interest.
If $\Delta{\vec v_{\rm sat}}\simeq v_{\rm sat}$, and $N \simeq 3\times10^3$ this
implies $v_{\rm ejection}\simeq 12\hbox{--}120 \kmps$.  
$N={\cal O}(3\times10^3)$ is a
reasonable number of corrections when it takes one or two days
to reposition BOSS
for a new occultation, since it 
implies a satellite lifetime of about $5$ years.
With careful scheduling, and clever orbital dynamics,
typical course corrections might not require 
$\Delta{\vec v_{\rm sat}}\simeq v_{\rm sat}$.
Off-the shelf, low-cost ion engines are currently available
with ejection velocities of $20 \kmps$, and
more expensive systems with $30 \kmps$ performance have been developed.

The issue of steerability therefore comes down to the sizes of the course
corrections in which we are interested, the time scale over which we need to
move the satellite to the new orbit, the effectiveness with which one
can use the solar-sailing technique,
and the state-of-the-art in high ejection velocity drives.
The smaller the corrections and the
longer we can wait, the less fuel we will need.  Since planets are not
transient sources, we can wait a relatively long time 
and use clever orbital dynamics
to do much of the work for us.  How many corrections we can make, therefore
depends on exactly how we use the satellite.  A reasonable program of
observations seems possible, though careful scheduling of observations and
clever design of orbital parameters will be essential.

\section{Diffraction}
\label{sect:diffract}

Because we are interested in observing systems with very high contrast,
we must include the effects of diffraction.
The angular width of the satellite diffraction pattern 
in the region far from the satellite is 
\begin{equation}
\label{diffraction}
\Delta\theta_{\rm diff}\simeq \left\{ 
  \begin{array}{ll}
    \frac\lambda{2R}, & \frac\lambda{2R} \ge \frac{2R}{\Dsat}\quad
    \hbox{(Fraunhoffer limit)} \\
    \sqrt{\frac\lambda{\Dsat}}, & \frac\lambda{2R} \le \frac{2R}{\Dsat}\quad
    \hbox{(Fresnel limit)} \\
  \end{array}
\right. .
\end{equation}
Here $2R\simeq 70\meter$ is some typical size of the occulting patch.
For $\lambda\simeq 1\micron$ and $\Dsat\simeq 1\times10^8\meter$,
the distance to the satellite, 
the transition from Fresnel to Fraunhoffer behavior occurs for 
$R\simeq \sqrt{\lambda \Dsat}/2 \simeq  5\meter$. 
For $R=5\meter$ and $\lambda=1\micron$,
$\theta_{\rm diff}\simeq 1\times 10^{-7} = 21\mas$.

Inspection of (\ref{diffraction}), shows that once the size of the satellite
has reached $R_{\rm crit}\simeq\sqrt{\lambda \Dsat}/2$ no further improvement in
resolution can be obtained just by increasing the size.  However, the fraction
of the intensity of the occulted source which is not blocked decreases with
the area of the occulter.  For this reason, we choose $R\gg 5\meter$.  The
price is that we must use the Fresnel diffraction pattern of the sources as
occulted by the satellite and as observed through the telescope.
Here we briefly outline the necessary steps.

The diffraction pattern for an arbitrary aperture ${\cal A}$ is given by
\be 
U = - \frac{Bi}{\lambda r's'} \int\!\!\int\nolimits_{\cal A} \tau(S) e^{ik
(r + s)}\,dS
\ee
where $B$ is a normalization constant related to the intensity
of the source, $dS$ is a surface element in the
aperture plane, $\tau(S)$ is the transmission function, $r$ ($s$) is
the distance between the observation (source) point and a point in the aperture
plane, and $r'$ ($s'$) is the distance between the observation (source) point
and the origin of the aperture plane.  Here we have assumed that the distances
between the aperture plane and the source and observation planes are large
and that all three planes are parallel.  We may expand $r+s$ in the
exponential in the limit of large separations.  Note that we must keep terms of
order $k\xi^2/2\RBOSS > 1$ where $\xi$ is a dimension in the aperture plane.
The presence of this higher order term means that we are working in the
Fresnel limit of diffraction.    A direct application of 
Babinet's principle  allows us to relate the diffraction pattern
due to the occulting satellite $\Umask (\rho,\phi)$,
to the complimentary problem of the diffraction
pattern of a finite circular aperture. Thus we can write $\Umask$ as
\be 
\Umask (\rho, \phi) = e^{-i\frac{k\tilde\rho^2}{2\Dsat}} + i\frac
k{\Dsat} \int_0^{\RBOSS} dr\,r e^{i\frac{kr^2}{2\Dsat}} J_0 \left(
\frac{k\tilde\rho r}{\Dsat} \right) \left[ 1 - \tau (r/\RBOSS) \right],
\ee
where $\tilde\rho^2 = \rho^2 + \rhosource^2 + 2 \rho\rhosource \cos
(\phi-\phisource)$, $(\rhosource,\phisource)$ is the position of the source
and $\tau$ is the transmission function of the satellite.
Throughout this discussion have assumed a circular occulter and an occulted
source on-axis.  The above
analysis has been performed for a square occulter and leads to comparable
results.

We want to look at this pattern with a telescope which also causes diffraction. 
The final pattern is given by an integral over the telescope
\be
U (w, \phi_0) = \frac{\sqrt\pi}{\lambda \Rtel } \int_0^{\Rtel} d\rho\,\rho
\int_0^{2\pi}d\phi\, 
e^{-ik\Rtel w \cos (\phi-\phi_0)} \Umask (\rho, \phi).
\label{Uimag}
\ee
Here $\Rtel$ is the radius of the telescope and $(w, \phi_0)$ is the angular
position (in radians) of the observation point $(x_0, y_0)$ with
respect to the center of the telescope.  The prefactor
normalizes the intensity in the observation such that 
\be 
\int_{0}^\infty dw\, w \int_0^{2\pi} d\phi_0\, \left| U (w, \phi_0)
\right|^2 = 1.
\ee

We have treated the telescope as a circular aperture with a CCD at infinity
which is used to image the diffraction pattern.  
The evaluation of $U$ is challenging.
Straightforward numerical integration of the highly oscillatory
integrand requires many hours to produce a single image.
Using a scheme in which $\Umask$ is expanded in 
Tchebyshev polynomials
we are able to evaluate all the integrals analytically
and express (\ref{Uimag}) as a sum over Bessel functions
which can be evaluated numerically.  A complete
image of an arbitrarily located  point source 
can thus be obtained in only a few minutes.

The transmission function allows us to apodise the diffraction pattern caused
by the satellite thus making the planetary region darker.  For the
transmission function we follow the suggestion of Hyde~(1998).
We write the transmission function as 
\be 
\tau_N (y) = \sum_{n=0}^N c_n y^n, 
\ee
where $N$ is the order of the occulter and we define $\tau_0=0$.  Here
\be
 y = \frac{\left (\frac r{\RBOSS}\right)^2 - \epsilon}{1-\epsilon}
\ee
and $\epsilon$ is the fractional radius of the center of the occulter that is
opaque ($\tau=0$).  
(All of the functions are made to satisfy $0\le \tau_N (y)\le1$ for $0\le y\le1$.)
Throughout this work we will focus on the fourth order occulter
\be
\tau_4 (y) = 35y^4 - 84y^3 + 70y^6 - 20y^7,
\ee
and consider $\epsilon=0.15$.
Note that this transmission
function is obtained by optimizing the on-axis nulling of the occulter.
Though this criterion is not optimal for planet detection, it provides a
simple function that improves performance and can be used as a 
starting point for future investigations.  

We should also note that real materials do not have perfect transmission.  In
the near IR it is possible to get 97\% transmission in a film with
an anti-reflective  coating. Thus we will replace $\tau$ with $(1-\delta)\tau$
and consider $\delta = 0.03$.
We will also use a constant transmission function of $1-\delta$ 
for the square in which our circular pattern is imprinted.
Smaller values of $\delta$, if achieved, 
could lead to improvement in the performance of the BOSS.

\section{The Satellite as a Source}

Just like a planet, BOSS would shine by reflection, 
both coherent and diffuse, and by thermal emission.
A serious concern is that BOSS might appear brighter 
than the sources  it is occulting and thus be of little value.
The concerns are less pronounced in the space-space configuration
than they are in the space-ground configuration
since by correctly orienting the satellite 
one can keep the sunlit side of the occulter away from the telescope.
However the situation is complicated by using a radially dependent
transmission function for the occulter since some sunlight is scattered
in and through the portions of the occulter which are not completely
opaque. Here we will discuss the constraints arising from 
all three sources: reflection, scattering, and thermal emission.

The flux of photons arriving from a star of bolometric magnitude
$m_b$ is approximately
\be
\label{eqn:phistar}
\Phi_{\star} = 
1.2 \times 10^{11} \left({T_\odot\over T_\star}\right) 10^{-0.4m_b} \pmsqs .
\ee
Since $m_b=-26.85$ for the Sun
\be
\Phi_{\odot} = 6.4\times 10^{21}\pmsqs .
\ee

We first consider the problem of reflection of sunlight off 
the satellite into the telescope.
Consider a reflective patch  on the satellite of area $A_{\rm patch}$
deliberately angled so as to reflect sunlight into the telescope.
This patch will reflect photons at a rate
\be
\left({d n_\gamma \over dt}\right)_{\rm reflect} = 6.4 \times 10^{21}
\frac{A_{\rm patch}}{\rm m^2}\unit{s^{-1}} .
\ee
{}From the position of the satellite however,
the Sun subtends a much larger solid angle of the sky than the telescope,
so only a fraction
\be
f_\Omega =  {(\Rtel/\DBOSS)^2 
                       \over (R_\odot/1.01\;{\rm AU})^2}
= 7.5\times 10^{-11} \left({\Rtel\over 4\meter}\right)^2
\left({\DBOSS\over 1\times10^8\meter}\right)^{-2}
\ee
of what is reflected will enter the telescope.
Here $\DBOSS$ is the distance from BOSS to the telescope and $\Rtel$ is the
radius of the telescope.
Thus the rate of photons collected by the telescope
from the reflecting patch would be
\be
\Gamma_{\rm reflected} = 
4.8\times 10^{11}\left ( \frac{A_{\rm patch}}{\rm m^2} \right)
\left({\Rtel\over 4\meter}\right)^2
\left({\DBOSS\over 1\times10^8\meter}\right)^{-2}\unit{s^{-1}} .
\ee
Compare this to the rate of photons received in the telescope from a star, 
\be
\Gamma_{\star} = 6 \times 10^{12-0.4m_b^\star} 
\left({T_\odot\over T_\star}\right) 
\left({\Rtel\over 4\meter}\right)^2 \unit{s^{-1}} .
\ee
We would like to keep the reflected light from being brighter than a 16
magnitude star,
as this is the level at which it will begin to interfere with the
performance of the occulter.  We see that this requires
\be
A_{\rm patch}\leq 5 \unit{mm}^2
\left({T_\odot\over T_\star}\right) 
\left({\DBOSS\over 1\times10^8\meter}\right)^2 .
\ee
This seems a strenuous requirement and implies that careful  attention
must be given to the shape of the exposed surfaces of the occulter,
especially the edges.
However, it is reassuring that the Sun subtends
only $6.8\times10^{-5}$ steradians, or $5.4\times10^{-6}$ of the sky,
and the telescope mirror subtends only $2.2\times10^{-13}$ steradians.
Random fluctuations of the occulter surface are therefore unlikely to reflect
light into the telescope.
So long as we are careful to appropriately orient the satellite 
and do not allow overly large fluctuations of the underlying surface
we should be able to meet this specification.   

Not all  of the sunlight incident on the satellite will be coherently reflected.
Some will be absorbed  and some will pass through the transmissive portions 
of the occulting body and be scattered diffusely.
Let $\cal S$ be the fraction of photons that are scattered diffusely.
Then, assuming isotropic scattering, the flux of  photons at the telescope 
resulting from diffuse reflection from the satellite will be
\be
\Phi_{\rm diffuse} =   \Phi_\odot {\cal S}
\left({A_{\rm BOSS} \over 4\pi \DBOSS^2}\right) .
\ee
For this to be less than the flux of a star of bolometric magnitude $m_b$
(cf.\ equation \ref{eqn:phistar})
\be
{\cal S} < 1.2\sci{-4} 
\left({\DBOSS \over 1\times 10^8\meter}\right)^2
\left({A_{\rm BOSS} \over (70\meter)^2}\right)^{-1}
\left({T_\odot\over T_\star}\right)
10^{-0.4(m_b^\star - 16)}.
\ee
The diffuse scattering of light in transparent films, such as polyimides,
which might be used for BOSS has been studied.  For a $10\micron$ film ${\cal
  S} < 10^{-4}$ at $1\micron$ is not unusual (e.g.~Kowalczyk, \etal~1994).

Intrinsic emission arises because BOSS is warmed by the sun.  
The integrated solar flux density at L2 is  approximately
\be
f_\odot \simeq 1.38\times 10^6 \unit{erg}\unit{cm^{-2}}\unit{s^{-1}} .
\ee
The power absorbed by the occulter is
\be P_{\rm abs} = \alpha f_\odot A_{\rm BOSS}\cos\theta, \ee
where $\alpha$ is the fraction of incident solar radiation energy absorbed,
and $A_{\rm BOSS}\cos\theta$ is the projected area  presented to the sun.
The power reemitted by BOSS is 
\be P_{\rm reemit} = e A'_{\rm BOSS} \sigma T_{\rm BOSS}^4. \ee
Here $e$ is the emissivity and $\sigma$ is the Stefan-Boltzmann constant.
Notice that $A'_{BOSS} \geq A_{BOSS}$,   since the satellite can
radiate from both sides, but absorbs sunlight from only  one.
If the satellite comes into thermal equilibrium with the solar radiation
then
\begin{eqnarray}
T_{\rm  BOSS} 
&=& \left( {\alpha f_\odot \over e \sigma} 
{ A_{\rm BOSS}\over A'_{\rm BOSS}} \cos\theta\right)^{1/4}, \nonumber\\
&=& 394\kelvin \left(\frac\alpha{e}{ A_{\rm BOSS}\over A'_{\rm BOSS}}
  \cos\theta\right)^{1/4}.
\end{eqnarray}
For emission from one side ($A'=A$) this gives  
$T_{\rm BOSS}=394\kelvin \left({\alpha\over e} \cos\theta\right)^{1/4}$,
while for equal two sided emission ($A'=2A$) it gives 
$T_{\rm BOSS}=332\kelvin \left({\alpha\over e}\cos\theta\right)^{1/4}$.
For many materials $\alpha/e\sim2$,
but typically during observations $\theta$ will  be greater than $60^o$,
so $\alpha \cos\theta/e$ should not be large.

The total photon flux at the telescope 
coming from the radiating satellite in thermal equilibrium
in a waveband from $\lambda_1$ to $\lambda_2$ is
\be
\Phi_{\rm thermal} = \alpha\cos\theta {A_{\rm BOSS}\over 2\pi \DBOSS^2} 
{f_\odot\over 2.7 k T_{\rm BOSS}} g(\lambda_1,\lambda_2,T_{\rm BOSS})  ,
\label{bossflux}
\ee
where
\be
g(\lambda_1,\lambda_2,T) = {1\over 2 \zeta(3)}\
\int_{hc/\lambda_1 k T}^{hc/\lambda_2 k T}
{x^2 dx \over e^x-1}
\ee
is the fraction of the total photons emitted in the relevant waveband.
Comparing this to the flux from an occulted star in the same waveband we find
\begin{eqnarray}
{\Phi_{\rm thermal}(\lambda_1,\lambda_2,T_{\rm BOSS}) \over
\Phi_{\rm occulted}(\lambda_1,\lambda_2,T_\star) } & = &
3\times 10^{5+0.4(m_b^\star-16)}\alpha\cos\theta
\left({T_\star\over T_\odot}\right)
\left({332\kelvin\over T_{\rm BOSS}}\right)
\left({A_{\rm BOSS}\over (70\meter)^2}\right) \nonumber \\
&& \times \left({1\times10^8\meter\over \DBOSS}\right)^2 
{g_{\rm thermal}(\lambda_1,\lambda_2,T_{\rm BOSS}) \over
g_{\rm thermal}(\lambda_1,\lambda_2,T_\star) } .
\end{eqnarray}
For $T=332\kelvin$,  and $T_\star=T_\odot=5778\kelvin$,
${\Phi_{\rm thermal}/\Phi_{\rm occulted}}$ varies from 
$1.4\times 10^{-8}$ for a $0.2\micron$ waveband centered 
at $1\micron$ 
to $2.8\times 10^{-3}$ at $1.5\micron$ 
to $1.7$ at $2.0\micron$.
At $T=394\kelvin$, these rise to 
$4.0\times 10^{-6}$ at $1\micron$,
$0.1$ at $1.5\micron$ 
and $28$ at $2.0\micron$.
Thus observations for wavelengths shortward of  about
$2.0\micron$ are unlikely to be seriously affected
by the thermal  emissions from BOSS.

Allowing the satellite to be as bright as a magnitude 16 solar type star
does not limit us to resolving pairs of magnitude 16 objects since only the 
shot noise in the flux from the satellite is relevant.
The ratio of the square root of detected satellite photons to the 
number of photons from a star of bolometric magnitude $m_b$
is 
\be {{\sqrt {N_{\rm BOSS}}} \over N_\star} 
=  3.2\times 10^{-10} 
\left[ {50 \meter^2\over A_{\rm telescope}} 
        {\second\over \Delta t}\right]^{1/2} 
{T_\star \over T_\odot} 10^{0.4m_b} .
\ee
For $T_\star=T_\odot$ and a $1\second$ exposure
(the typical time it takes a satellite orbiting L2 
to cross $0.1\mas$)
a 23.7 bolometric magnitude target star would give
$ {{\sqrt {N_{\rm BOSS}}} \over N_\star}  \simeq 1$.
It thus should be relatively easy to differentiate
between stars of magnitude 20 or less at separations
of better than $0.1\mas$.

\section{Planet Searches}
\label{sect:planet}

\begin{figure}
  \leavevmode\center{\rotate[r]{\epsfig{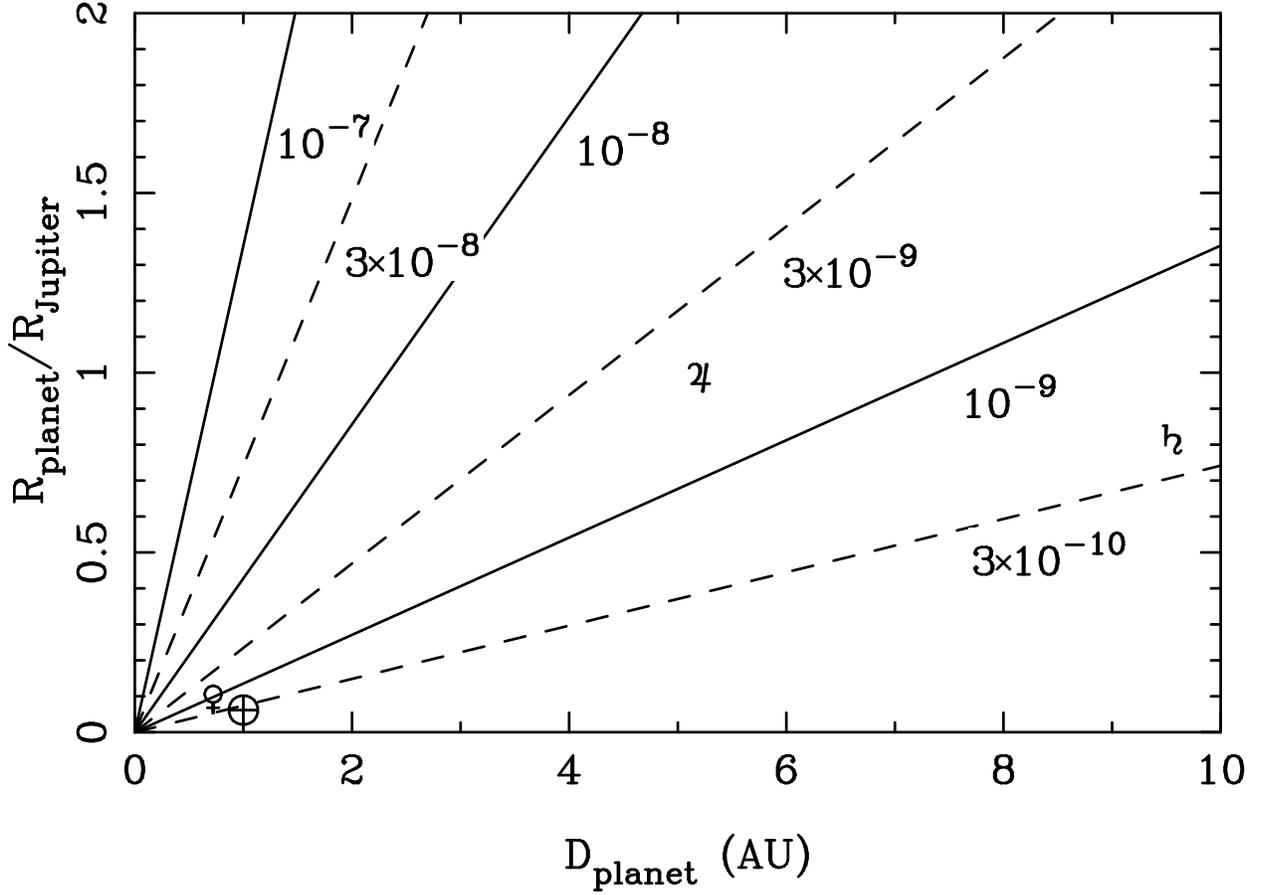}}}
  \caption{The detection capability of a planet with BOSS as a function of a
    planet's size and distance from the star.  The curves show the relationship
    between the size of a planet and its distance from a star for relativity
    intensities of $10^{-7}$, $3\sci{-8}$, $10^{-8}$, $3\sci{-9}$, $10^{-9}$,
    and $3\sci{-10}$.  Here we have assumed the factor combining the albedo,
    phase angle, and illuminated fraction of Jupiter is equal to one.
    For a given detectable relative intensity (for example
    $3\sci{-9}$) all planets above this curve would be detectable. 
    Also shown in this figure are the brightest planets in our solar system
    (from left to right using their standard symbols), Venus, Earth, Jupiter,
    and Saturn.  We have plotted the planets using an effective radius that
    includes their albedo relative to Jupiter so that the 
    relative intensity curves
    are scaled the same for all the planet.
    }
  \label{fig:planet-reflect}
\end{figure}

\begin{figure}
  \leavevmode\center{\epsfig{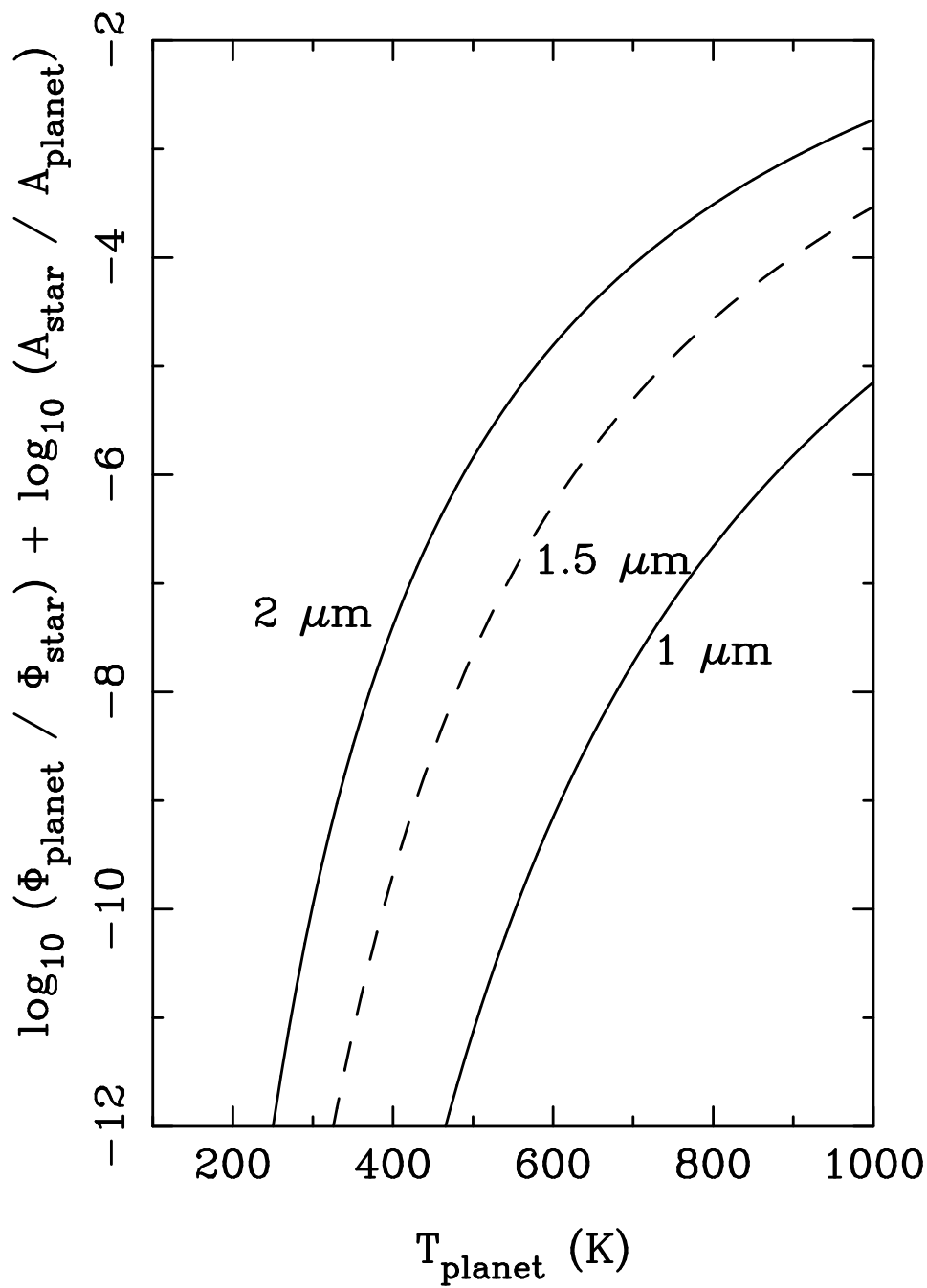}}
  \caption{The ratio of blackbody emission from a planet and a Sun-like
    ($T_{\rm star} = T_\odot = 5778\kelvin$) star.  Notice the extremely sharp
    fall off at low temperatures.  This prevents emission from being an
    important contribution to the intensity of detectable planets.  For
    Jupiter $\log_{10} \left (A_\odot/A_{\rm Jupiter} \right) = 2$ and $T_{\rm
      Jupiter} = 125\kelvin$.}
  \label{fig:planet-emit}
\end{figure}

Planets shine in two ways, in reflected light and in emitted light.  In
reflected light, the brightness of a Jupiter-like planet orbiting $1\AU$ from
a star is approximately $10^{-7}$ times that of the star; falling off as
$1/r^2$ as the orbital radius increases, but growing as the radius of planet
squared (see figure~\ref{fig:planet-reflect}).  In emitted light, a planet
glows as approximately a black body characterized by its temperature (see
figure~\ref{fig:planet-emit}), though molecular absorption can alter that
dramatically in certain wavebands.  A planet's temperature is a strong
function of its age, the central star's type and proximity, 
atmospheric composition, and internal heat sources;
however, typical brightness ratios are still $\ltsim
10^{-8}$ except for very young gas giants or planets very close to the central
star (see e.g., Burrows \etal~1997).  We will focus on a 1 micron wave band
(0.9--1.1\micron) for our studies since the diffraction from an 8-m telescope
is not too severe in this waveband.

To study the benefits of employing BOSS in conjunction with a space telescope
we have simulated the images produced by our solar system at $3\parsec$ and
$10\parsec$ occulted by BOSS and
viewed by a 8-m telescope.  For our solar system we have included the Sun
occulted on-axis and the four brightest planets; Venus, Earth, Jupiter, and
Saturn.  We have used the BOSS geometry
described in section~\ref{sect:geometry} and have assumed that BOSS is located
$1\times 10^8\meter$ from an 8-m space telescope at L2.  All images have been
calculated for the 1\micron\ waveband.
Images of the occulted star and unocculted planets are calculated for 3
different wavelengths in this waveband as described in
section~\ref{sect:diffract}.  These images are then smoothed with the point
spread function~(PSF) of the telescope.  The PSF we employ has a diffraction
limited core and a Gaussian halo with a $1\as$ FWHM that contains $10^{-3}$ of
the light.  The smoothed images provide the templates of the
occulted star and unocculted planets.  From these templates we can construct
an image of a solar system.  To construct a solar system image we first scale
the planets' intensities by a fraction of the unocculted star's intensity.  We 
then include a Gaussian random component equal to 5\% the intensity of each
pixel to account for uncertainties in the PSF\@.  After adding the planets to
the occulted star we Poisson sample the resulting image to produce an image of 
the solar system.  To identify planets, the template of the occulted star is
subtracted off leaving behind the planets and the noise.  A planet is
detectable if it can be identified above this noise.

\placetable{tab:solarsystem}

\begin{figure}
  \leavevmode\center{\epsfig{figure=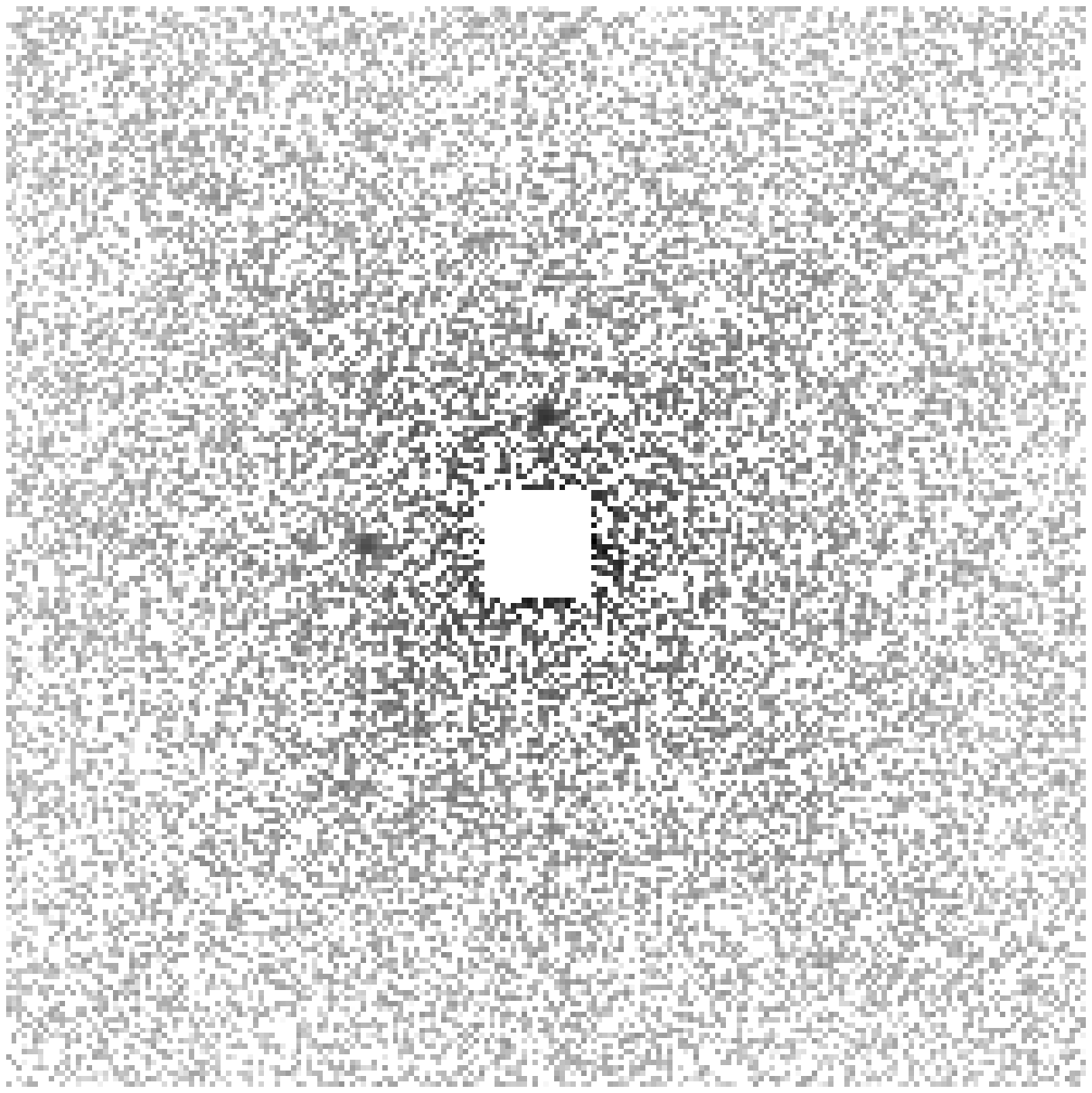,width=5in}}
  \caption{A $2\as$ by $2\as$ CCD image of the log of the intensity for our 
    our solar system at $3\parsec$ from BOSS as observed in the $1\micron$
    waveband by an 8-m space telescope in a $3000\second$ exposure.
    The Sun at this distance has a bolometric magnitude of $m^\odot_b = 2.1$
    and is located at the center of the satellite.  
    The central $0.2\as$ square (roughly the size of BOSS) has been cut out
    of the image.
    Both Venus (above) and the
    Earth (left) are observable.  Jupiter and Saturn are easily
    detectable but are outside the field of view of this image.}
  \label{fig:solar3}
\end{figure}

\begin{figure}
  \leavevmode\center{\epsfig{figure=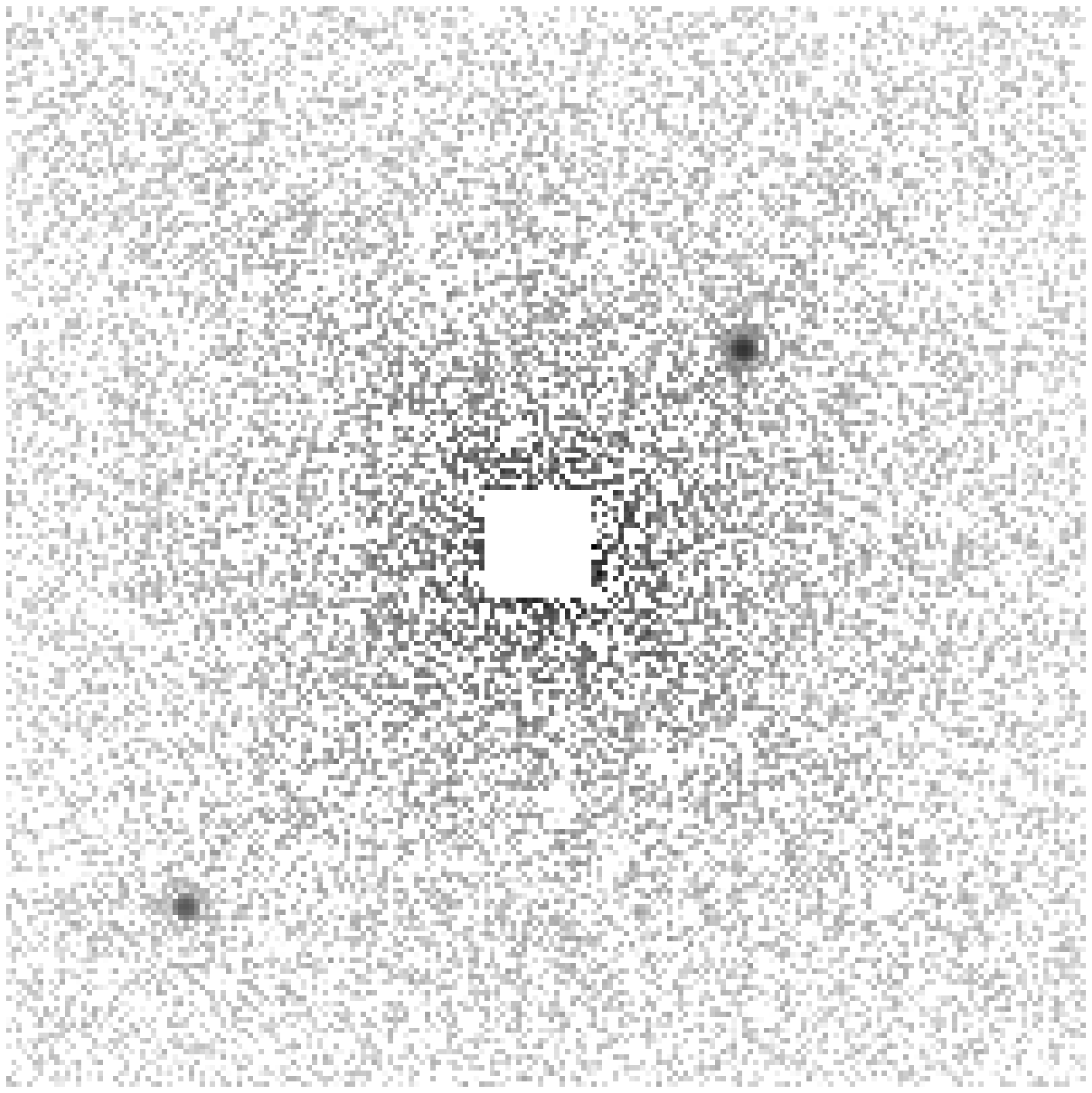,width=5in}}
  \caption{A $2\as$ by $2\as$ CCD image of the log of the intensity for our 
    our solar system at $10\parsec$ from BOSS as observed in the $1\micron$
    waveband by an 8-m space telescope in a $3000\second$ exposure.
    The Sun at this distance has a bolometric magnitude of $m^\odot_b = 4.72$
    and is located at the center of the satellite.  
    The central $0.2\as$ square (roughly the size of BOSS) has been cut out
    of the image.
    Both Venus and the
    Earth are occulted by BOSS\@.  Jupiter (upper right) and Saturn (lower left)
    are easily detectable.}
  \label{fig:solar10}
\end{figure}

Figures~\ref{fig:solar3} and~\ref{fig:solar10} show our solar system
$3\parsec$ and 
$10\parsec$ away from us, respectively.  The parameters describing out solar
system are given in table~\ref{tab:solarsystem}.  The bolometric magnitude of
the Sun at $3\parsec$ is $m^\odot_b = 2.1$ and at $10\parsec$ is $m^\odot_b =
4.72$.  The images are $2\as\times2\as$ CCD images with
$0.01\as\times0.01\as$ pixels.  The central $0.2\as$ square
(roughly the size of BOSS) has been cut out.  At $3\parsec$ Venus and the
Earth are visible by-eye (see figure~\ref{fig:solar3}).  Both Jupiter
and Saturn would be easily observable but are outside the CCD shown here.  At
$10\parsec$ Venus and the Earth are occulted by BOSS but both Jupiter and
Saturn are easily identified (see figure~\ref{fig:solar10}).  A wider variety
of images in color can be found at http://erebus.phys.cwru.edu/$\sim$boss/.

In figures~\ref{fig:solar3} and~\ref{fig:solar10} our solar system is shown in 
reflected light.  The relative intensity for Jupiter was calculated using its
Bond albedo and assuming uniform scattering into $2\pi$ steradians.  The
intensity for Jupiter quoted in table~\ref{tab:solarsystem} is based on this.
The intensity for the rest of the planets quoted in
table~\ref{tab:solarsystem} are scaled from Jupiter using their size, distance,
and albedo relative to Jupiter.  We should also note that many improvements can 
be made in the analysis described above.  A statistical image reconstruction
technique could more reliably detect planets and may allow for identification
of planets that the ``by-eye'' technique misses.  Also, the reflected light
from a  planet would be polarized whereas starlight is not.  Thus even further
suppression of the starlight should be possible.  Finally we comment that the
$1\micron$ waveband is a fairly good choice 
for gas giant planets.  Although $\rm CH_4$ (which is prevalent in gas giant
atmospheres) has an absorption feature at $1\micron$, it is fairly weak and
sufficient flux comes from the two edges of the waveband to compensate for
this.  Above about $1.1\micron$ strong $\rm CH_4$ absorption dominates the
spectrum and drastically reduces the intensity of reflected light.

Blackbody emission at $1\micron$ from planets similar to those in our solar is 
not important since the planets are too cool.  A Jupiter-like planet would
only be detectable at $1\micron$ if
it were extremely hot ($T_{\rm planet} > 800\kelvin$, see
figure~\ref{fig:planet-emit}).  Notice the sharp increase in
intensity shown in figure~\ref{fig:planet-emit} at higher
wavelengths.  In particular at $2\micron$ planets down to a temperature of
$T_{\rm planet} \approx 400\kelvin$ could be observed in emission if a similar
relative intensity limit can be achieved.  However Jupiter has an
effective blackbody temperature of $T_{\rm Jupiter} = 125\kelvin$ and would
still not be observed in emission at $2\micron$. In fact, about $5.5\micron$
is where Jupiter starts radiating brightly.  Thus emission will not
likely play an important role at $1\micron$ for this study.
Larger wavelengths require larger
telescopes to cut down on the size of the diffracted images.  Even at
$1.5\micron$, the ``by-eye'' identification technique employed here is not
sufficient for finding faint planets since the light of both the planet and
occulted star are spread out over larger regions.

\section{Conclusion}

Occultations provide a powerful technique for resolving both small angular
scales and large separations in intensity.  BOSS in a space-space
configuration will offer distinct improvements over a lone space telescope
allowing it to not only detect planets but also directly image them.  BOSS
blocks all but $4\sci{-5}$ of 
the light from a star in the $1\micron$ waveband.  Furthermore, by doing the
occultation well outside the telescope, degradation due to scattering inside
the optics is minimized.  The low effective gravity at L2 offers the
possibility of obtaining very long exposures.

In this work we have shown that a $70\meter\times 70\meter$ BOSS in
conjunction with an 8-m space telescope at L2 would allow for the direct
imaging of Venus, Earth, Jupiter, and Saturn at $3\parsec$.  
Earth and Venus would remain visible out to at least 5 pc.
Both Jupiter and Saturn would be visible at $10\parsec$ 
and remain so out to about $20\parsec$.  
Solar systems with Jupiter-like planets closer to their star
would be easily observable.  
Furthermore, photometric and spectrographic
observations could be considered for these objects.

Many improvements can be made over the techniques employed here to aid in
finding dim planets.  A sophisticated reconstruction technique would greatly
improve the ``by-eye'' technique presented here.  Also using polarized light
from the planet versus the unpolarized light from the star would significantly
increase the separation of planets from the star.  In spite of the simple
techniques used here BOSS has been shown to be a powerful tool for discovering
and imaging planets.

We have discussed only briefly high resolution observations using BOSS,
though our early estimates lead us to anticipate  a capability
to resolve binary sources a the well sub-milliarcsecond level.
Further discussions will appear in upcoming papers, but BOSS's
potential for high resolution imaging,
looks to be as exciting  as its promise for planet discovery.

\acknowledgements
The authors would like particularly to thank 
C. Beichman for very extensive conversations on AO, planet finding,
and IR astronomy, for ongoing encouragement and for arranging
financial support from JPL;
Art Chmielewski for financial and other support and, 
with Chuck Garner and Rick Helms,
for educating them about inflatables;
L. Close for telling them about state-of-the-art in AO;
M. Dragovan  for many useful comments and suggestions;
F. Kustas for mission and systems analysis;
L. Lichodziejewski for extensive design efforts;
H. Morrison and P. Harding for educating them
about PSF's, filters, and various nitty gritty astronomical details;
and V. Slabinski for lengthy discussions on orbital dynamics
and satellite control capabilities.
They would also  like to thank
B. Madore for his interest and encouragement;
P. McGregor for information on IR backgrounds;
P. Taylor for assistance with questions on optics; 
W. Tobocman for special functions;
and A. de Laix, L. Krauss, H. Mathur, M. Trodden and T. Vachaspati  for 
ongoing useful comments.
GDS would like to thank C. Alcock for 
very helpful discussions and encouragement
during the initial phase of this research, and
G. Marcy, B. Matisack, S. Drell and S. Tremaine for useful input.  
This work was supported by 
a CAREER grant to GDS from the National Science Foundation,
a DOE grant to the theoretical particle and astrophysics group at CWRU, 
by a grant from NASA's Jet Propulsion Laboratory,
and by funds from CWRU.
Allan Fetters did some of the early detailed analysis of integration times
for the space-ground configuration and prepared the BOSS home page.  
David Rear contributed heavily to early programming in the analysis of 
imaging capabilities.

\clearpage

\begin{deluxetable}{lcccccc}
\tablecolumns{7}
\tablewidth{0pt}
\tablecaption{Parameters for our Solar System\label{tab:solarsystem}}

\tablehead{
\colhead{} & \colhead{} & \colhead{} & \colhead{} & \colhead{} & 
 \multicolumn{2}{c}{Angular Separation (as)} \\
\cline{6-7} \\
\colhead{Planet} & \colhead{Radius ($R_{\rm Jup}$)} & \colhead{Distance (AU)}
& \colhead{Albedo} & \colhead{Flux ($\Phi_\odot$)} & \colhead{$3\parsec$}
& \colhead{$10\parsec$}  } 
\startdata
Venus   & 0.087 & 0.72 & 0.76 & $1.2\sci{-9}$  & 0.24 & 0.07 \nl
Earth   & 0.091 & 1.00 & 0.39 & $3.4\sci{-10}$ & 0.33 & 0.10 \nl
Jupiter & 1.000 & 5.20 & 0.51 & $2.0\sci{-9}$  & 1.73 & 0.52 \nl
Saturn  & 0.837 & 9.56 & 0.50 & $4.0\sci{-10}$ & 3.19 & 0.95 \nl
\enddata
\tablecomments{For each planet the radius is given in units of Jupiter's
  radius, the distance from the Sun is given in AU, and the flux of reflected
  light is given relative to the flux of the Sun.  The angular separations are 
  given for the solar system at $3\parsec$ and $10\parsec$ from BOSS\@.}
\end{deluxetable}

\end{document}